# How are Identifiers Named in Open Source Software? On Popularity and Consistency

Yanqing Wang[a*], Chong Wang[b], Xiaojie Li[a], Sijing Yun[a], Minjing Song[a]
[a]School of Management, Harbin Institute of Technology, Harbin 150001, China
[b]Huaihai Institute of Technology, Lianyungang 222005, China

*Abstract*—**With the rapid increasing of software project size and maintenance cost, adherence to coding standards especially by managing identifier naming, is attracting a pressing concern from both computer science educators and software managers. Software developers mainly use identifier names to represent the knowledge recorded in source code. However, the popularity and adoption consistency of identifier naming conventions have not been revealed yet in this field. Taking forty-eight popular open source projects written in three top-ranking programming languages Java, C and C++ as examples, an identifier extraction tool based on regular expression matching is developed. In the subsequent investigation, some interesting findings are obtained. For the identifier naming popularity, it is found that Camel and Pascal naming conventions are leading the road while Hungarian notation is vanishing. For the identifier naming consistency, we have found that the projects written in Java have a much better performance than those written in C and C++. Finally, academia and software industry are urged to adopt the most popular naming conventions consistently in their practices so as to lead the identifier naming to a standard, unified and high-quality road.**

*Keywords--- identifier naming convention, coding standards, open source project, identifier naming popularity, identifier naming consistency*

## I. INTRODUCTION

In rapid-growing software industry, more and more attention has been drawn to software maintainability and ultimate quality assurance [1,2]. Product reliability and maintainability has become the spotlight of software industry. The viewpoint that higher readability and understandability are helpful for software maintenance has been generally accepted [3,4,5]. Some popular integrated development environments (IDEs), such as *Netbeans*, *Eclipse* and *Code::blocks*, have the automatic formatting function to beautify indentation or layout of source code, but they cannot beautify identifier naming because identifier is more like an artifact than a machine-generated character sequence. Identifier name is a major way for programmers to communicate concepts [4] and represent the knowledge recorded in source code [6].

Lots of researchers have discovered that identifier naming conventions are of paramount importance for software quality. In computer programming, a naming convention is a set of rules for choosing the character sequence to be used for identifiers which denote *variables*, *types*, *objects* and *functions* etc. in source code and documentation. Identifiers occupy a large percentage of source code in programming, more than two thirds in some projects and identifier naming has a close relationship with the quality and comprehensibility of a software system [4]. Butler et al have proved the negative effect between identifier naming defects and source code reliability from statistical view by using *findbugs* [7]. Some empirical studies and dynamic feedback have verified the positive effect between identifier naming and source code readability [4,6,8].

Fortunately, academia and software industry have proposed several identifier naming rules. Carter put forward an identifier naming criterion to enhance software maintainability [3]. Programmers should reach an agreement on all identifiers' abbreviation which may possibly be used before writing programs. Although being not easy to be popularized in today's source code writing, it gave us a new idea to establish appropriate identifier naming standard. In last decades, such software organizations as *IBM*, *Microsoft*, *Bell Laboratory* made great contribution to coding standards including identifier naming. Many scholars have proposed their own identifier naming rules as well [9,10,11]. Some of the rules are language-specific, such as *Java* identifier naming guidelines [11] and *C* identifier naming guidelines [10,12].

---

[*] Corresponding author
*E-mail addresses*: yanqing@hit.edu.cn (Y.Q. Wang), zgwangc@sina.cn (C. Wang)

So far, the research on identifier naming rules has been focused on three aspects: (1) to enhance understandability of identifier naming, such as compiling an identifier dictionary [13,14], suggesting all the identifiers be composed by the type names [14,15] and limiting length of identifiers [16]; (2) to analyze the identifier names according to syntax rules [10,13,17]; (3) to dig the identifiers' constituent from the semantic angle [4,12,18]; (4) to investigate the impact of program identifier style on human comprehension [19].

However, there are still some major issues on identifier naming nowadays, which result from many aspects such as programmers' cultural practices or educational issues [8]. First, the choice of naming conventions (and the extent to which they are enforced) is often an enormously controversial issue, with partisans holding their viewpoint to be the best and others to be inferior. Second, even with known and well-defined naming conventions in place, some organizations may fail to consistently adhere to them, causing inconsistency and confusion of identifier naming even in a single project. These challenges may be exacerbated if the naming convention rules are internally inconsistent, arbitrary, difficult to remember, or otherwise perceived as more burdensome than beneficial. Therefore, the research on popularity and consistency of identifier naming becomes more and more urgent.

The remainder of the paper outlines as follows. Some necessary background information is presented in Section 2. Section 3 describes the investigation design. The results are analyzed in Section 4. Section 5 remarks this paper and presents some suggestions to software educators and software industry.

## II. BACKGROUNDS

TIOBE Software, a world-level coding standards company, presents the popularity ranking of programming languages every month. The update data are illustrated in Appendix A [20]. Thus, we focus our study on the top three popular programming languages including *C*, *C++* and *Java*.

### A. The identifier naming conventions we focus in this study

To the best of our knowledge, five naming conventions *Hungarian*, *Camel*, *Pascal*, *Underline* and *Capital* are widely adopted.

*(1) Hungarian notation.* One or more lowercase letters are used as the prefix of an identifier, so as to identify the scope and type of the identifier. After the prefix is one or more words with first letter uppercase, and the word should indicate the purpose of the variable [21,22]. The prefixes of *Hungarian* naming convention are listed in Appendix B. An identifier example in *Hungarian* naming convention is:
  *int lpQueueHead ;*

*(2) Camel naming convention.* It is also spelled *camel case*. Uppercase letters are taken as word separators, lowercase for the rest. The initial letter of the first word is in lower case as a camel is bowing its head. For example,
  *printEmployeePaychecks( ) ;*

*(3) Pascal naming convention.* It is similar to the *Camel* naming convention except that the initial letter of the first word is uppercase. For example:
  *public class BankAccount {*
    *...*
  *};*
  *new Circle(1.0) ;*

*(4) Capital naming convention.* A naming convention in which any identifier is composed of one or more words (separated with underline mark) written totally in uppercase letters, e.g. *NUMBER_OF_STUDENTS*. Most of time, this naming convention is used to define constants.

*(5) Underline naming convention.* Similar to *Pascal* and *Camel*, identifier in this convention is composed of one or more words or acronyms except that the separator between words is not an uppercase letter but an underline mark. For example:
  *int student_account_no = 30 ;*
  *float monthly_interest = 4.5 ;*

Thought an identifier in *Capital naming convention* may have one or more underline marks in it, it belongs to the above category *Capital naming convention* but not *Underline naming convention*.

In addition, some naming conventions such as *Positional Notation* in COBOL and *Composite word scheme* in IBM's OF language are used in very specific developing environments (JCL, MS-DOS or IBM company) and their usages are evanishing. Similarly, *Pre-underline* naming convention, in which every identifier begins with one or more underline marks, is generally used to distinguish several similar identifiers by some *C* or *C++* programmers in very early years. Therefore, the relevant researches on them are not covered in this paper.

### B. RegEx-based identifier recognition

Some software such as *yacc* and *srcml* can extract identifier efficiently. However, to get more detailed information for the following analysis, we develop an algorithm to recognize identifier automatically. After an identifier has been extracted, regular expression is applied to match it with the five considered naming conventions. When extracting identifiers from a source program, it is not a good idea to scan the file word by word because the scanning process is often disturbed by non-identifiers, and each variable has its individual definition. The appearance times and line numbers cannot be obtained only by addressing definition sentences so that two arrays are used. One array carries the identifiers obtained from definition sentences, the other stores the identifiers which are extracted when scanning the whole file and ignoring *punctuations*, *constants*, *reserved words*, etc. The latter array also helps system get occurrence times and line numbers of a certain identifier.

After being extracted, each identifier is attempted to match with pre-constructed regular expressions, as listed in Table 1, so that its naming convention will be determined. As mentioned above, five categories are defined and miscellaneous identifiers are ignored. By the way, two naming conventions *Underline* and *Capital* have somewhat intersection, but it does not affect our conclusion in following sections because *Capital* naming convention occupies a quite small proportion.

**Table 1** Regular expressions of 5 naming conventions

| Naming Convention | *Regular Expression* |
|---|---|
| Camel | \b(([a-z]+([A-Z][a-z]*)+)|[a-z]{2,})\d*\b |
| Pascal | \b([A-Z][a-z]+)+\d*\b |
| Underline | \b(([a-z]+(\d*)+_)+([a-z]*\d*)+)\b |
| Hungarian | \b([gmcs]_)?(p|fn|v|h|l|b|f|dw|sz|n|d|c|ch|i|by|w|r|u)(Max|Min|Init|T|Src|Dest)?([A-Z][a-z]+)+\d*\b |
| Capital | \b([A-Z]*(\d*)+_)*([A-Z]*\d*)+\b |

C. *2.3 The tool*

In computer science and software engineering research, experimental systems and toolkits play a crucial role [23]. To undertake our study well, we design and develop an application tool in Java with the IDE *netbeans*, whose human computer interface is illustrated in Fig. 1. The tool has three features as follows.

(1) This tool can deal with three kinds of popular languages *C*, *C++* and *Java*. Experiment operator can choose project language by clicking one of the three "Language" radio buttons. Some open source projects are written in hybrid languages, for example, *OpenOffice* is written in both *C++* and *Java*. Thus, the language option is designed to make sure the tool just focuses on specific file types (extension name) and ignores others. The considered files types of three languages are shown in Appendix C.

(2) This tool can run in three different modes: *single file* mode, *single project* mode and *multiple project* mode, as illustrated in Fig 1. In *single project* mode, when one folder name is selected, all files in it will be scanned fully recursively and the files with satisfactory types will be dealt with. In *multiple project* mode, the selected folder name is not the project name but a set of projects, in which each sub-folder is a project name.

(3) There are three text areas which are used to hold *project names*, *file names* and *detailed or summary results*. The checkbox buttons *Silent Mode* and *Debug* are used to control the granularity of output information. When *Silent Mode* is on, no detailed information is output to console. When *Debug* is on, some information which helps find defects will be shown in result area.

After each project has been scanned, the summary result will be stored into one MS SQL table, as illustrated in Appendix D, as one record row.

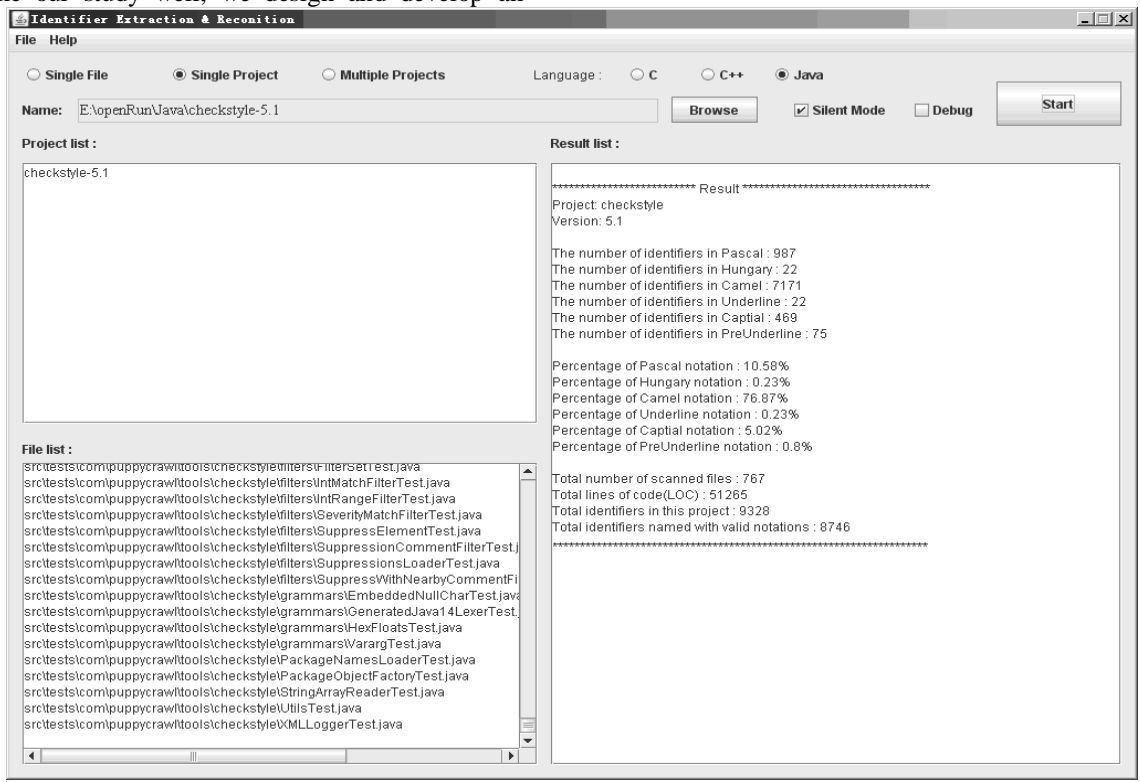

**Fig. 1** User interface of the tool for extracting and recognizing identifiers

## III. INVESTIGATION DESIGN

### A. Selection of open source projects

Open source projects are determined as our investigation data source. The analysis of open-source code is drawing much attention from academia [24]. Although Lawrie et al have proved that open and proprietary source include different identifier quality [14], it does not matter to select open source projects as the data source since identifier quality is not our concern in this research. Finally, 10 projects in *C* (12.4 MLOC), 14 projects in *C++* (17.7 MLOC) and 24 projects in *Java* (7.1 MLOC), totally 37.2 MLOC approximately, are involved in this study. Some leading open source organizations such as *Sourceforge*, *sun*, *GNU* and *apache* own the licenses of majority of the selected projects.

### B. Data preparation

With the developed tool just mentioned, the 48 selected projects are input into the system grouped by programming languages and the output are stored in the pre-defined database table. Besides *project name*, *version* and the identifier number in the five considered naming conventions mentioned above, the total number of valid identifiers and total LOC are also recorded. The naming convention usages in 3 languages are shown in Appendix E through G.

### C. Statistical method

In probability theory and statistics, the *coefficient of variation* (*CV*) is a normalized measure of dispersion of a probability distribution. It is defined as the ratio of the standard deviation to the mean:

$$CV = s/\bar{x} \qquad \text{E.1}$$

where *CV* stands for *coefficient of variation*, *s* denotes *standard deviation*, $\bar{x}$ is the *mean* of all parameters. The greater *CV* is, the more dispersion variation degree of the data set has.

In this paper, *CV* is applied to compare the dispersion degree of identifiers writing in all kinds of naming conventions in open source projects. By computing the dispersion index, we can judge whether the using of naming conventions is centralized or not. A greater *CV* means a higher fluctuation and a higher consistency of naming conventions, i.e. only a few naming conventions are being adopted in one project. Afterwards, the investigation is undertaken with the help of professional statistical tools such as SPSS 16.0 and MS Excel 2003.

## IV. DATA ANALYSIS

In order to prove that our research is based on a feasible classification of identifier naming conventions and the considered five naming conventions are sufficient for this study, the data in Appendix E through G are summarized group by programming languages (*C*, *C++*, *Java*), as depicted in Table 2. An index *match ratio* is defined to mean the number of matched identifiers (five considered naming conventions) over the total number of all identifiers.

The data in Table 2 shows that, the identifiers in five naming conventions account for a large ratio in the total identifiers. The *match ratios* of *C*, *C++* and *Java* projects are 88.71%, 83.88% and 91.87% correspondingly. It can be concluded that these five naming conventions we focus are widely adopted among software programmers and organizations.

### A. Which naming conventions have higher popularity?

In order to comprehend the popularity of the five naming conventions, their distributions (the number of each identifier over Total identifier in percentage) are listed in Table 3.

For the investigation of popularity, from the total distribution of five naming conventions, no matter in *C*, *C++* or *Java*, *Camel* is the most popular convention, which accounts for 46.99% in *C*, 51.13% in *C++*, and 85.39% in *Java*. *Underline* convention occupies 44.16% in *C*, almost as much as *Camel*. In *C++*, the ratio of *Hungarian* convention is far greater than *C* and *Java*. However, the ratio of *Capital* convention is almost the same in *C*, *C++* and *Java*. Thus, we conclude that *the five naming conventions have different popularity*.

Furthermore, from Table 3 and Fig. 2, it is easy to find that the popularities of naming conventions in projects written in different programming languages vary. The popularities of five naming conventions in *C, C++* and *Java* projects are ordered as in Table 4. Therefore, we know that *the popularity does relate with programming languages*.

**Table 2** Total numbers of matched identifiers and total match ratio of 3 programming languages

| *Language* | *C* | *C++* | *Java* |
|---|---|---|---|
| Pascal | 29,710 | 221,389 | 78,837 |
| Camel | 535,132 | 760,603 | 1,026,215 |
| Hungarian | 5,072 | 298,335 | 6,351 |
| Underline | 502,930 | 137,225 | 5,180 |
| Capital | 46,661 | 60,682 | 72,311 |
| *matched* | *1,119,505* | *1,478,234* | *1,188,894* |
| *total* | *1,261,940* | *1,762,233* | *1,294,165* |

| | match ratio | 88.71% | 83.88% | 91.87% |

**Table 3** Distribution of 5 naming conventions in 3 programming languages

| Language | Pascal | Camel | Hungarian | Underline | Capital |
|---|---|---|---|---|---|
| C | 2.61% | 46.99% | 0.45% | 44.16% | 4.10% |
| C++ | 14.88% | 51.13% | 20.05% | 9.22% | 4.08% |
| Java | 6.56% | 85.39% | 0.53% | 0.43% | 6.02% |

**Table 4** Popularity order of 5 naming conventions in 3 programming languages

| Language | 1st | 2nd | 3rd | 4th | 5th |
|---|---|---|---|---|---|
| C | Camel | Underline | Capital | Pascal | Hungarian |
| C++ | Camel | Hungarian | Pascal | Underline | Capital |
| Java | Camel | Pascal | Capital | Hungarian | Underline |

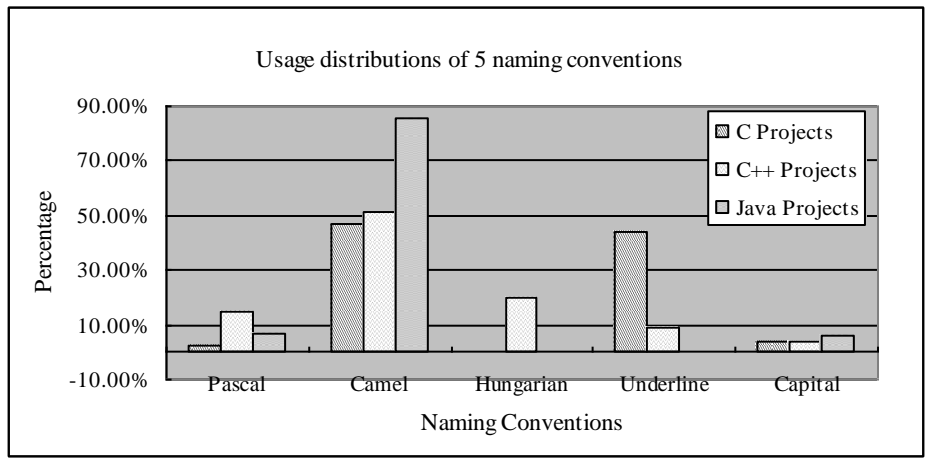

**Fig. 2** Usage distribution of five naming conventions grouped by programming languages

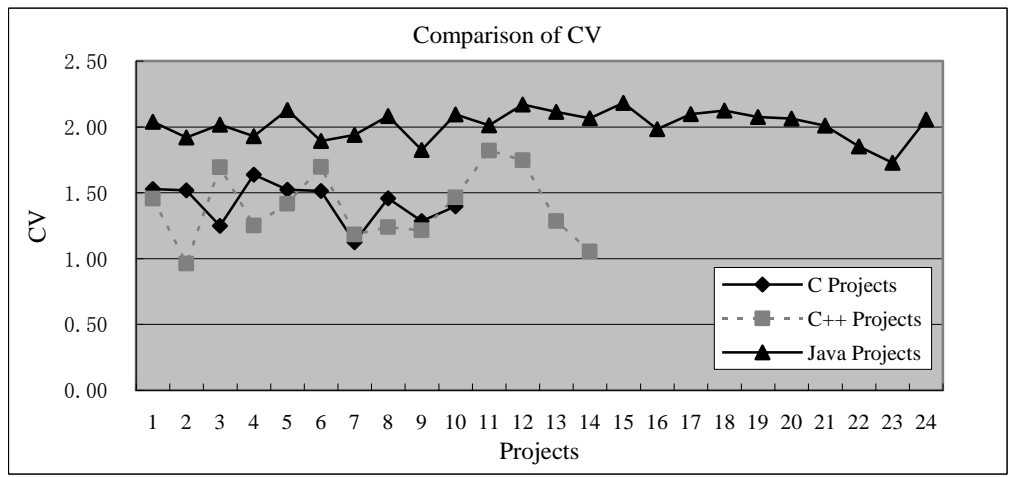

**Fig. 3** Comparison of CV in three programming languages

*B. In which languages, projects have higher naming consistency?*

For each project, the *CV* of naming conventions has been calculated. Through comparing every *CV* with each other, we can obtain the evaluation of dispersion degree of naming conventions. The conclusion is depicted as in Table 5. The corresponding line graph, as illustrated in Fig. 3, is based on the data in Table 5.

**Table 5** Coefficients of variation in all projects

| No. | C Project | C CV | C++ Project | C++ CV | Java Project | Java CV |
|---|---|---|---|---|---|---|
| 1 | haproxy | 1.53 | UnitTest | 1.45 | tiles | 2.04 |
| 2 | cherokee | 1.52 | sockets | 0.96 | ftpserver | 1.92 |
| 3 | zabbix | 1.25 | GigaBASE | 1.69 | mina | 2.02 |
| 4 | nginx | 1.64 | ekiga | 1.25 | velocity | 1.93 |
| 5 | httpd | 1.52 | LiteSQL | 1.42 | FreeMind | 2.13 |
| 6 | net-snmp | 1.51 | PorkerTH | 1.70 | freemarker | 1.89 |
| 7 | subversion | 1.12 | Nestopia | 1.18 | checkstyle | 1.94 |
| 8 | MingGW | 1.46 | codeblocks | 1.24 | Dependency | 2.08 |
| 9 | gcc | 1.28 | mysql | 1.21 | pmd | 1.82 |
| 10 | linux | 1.40 | CLucene | 1.46 | struts | 2.09 |
| 11 | | | qt | 1.82 | dwr | 2.01 |
| 12 | | | firefox | 1.75 | hibernate | 2.17 |
| 13 | | | CodeLite | 1.28 | findbugs | 2.11 |
| 14 | | | Shareaza | 1.05 | ant | 2.06 |
| 15 | | | | | JFreeChart | 2.18 |
| 16 | | | | | jruby | 1.98 |
| 17 | | | | | tomcat | 2.10 |
| 18 | | | | | JasperReports | 2.12 |
| 19 | | | | | axis2 | 2.07 |
| 20 | | | | | spring | 2.06 |
| 21 | | | | | openjpa | 2.01 |
| 22 | | | | | JEdit | 1.85 |
| 23 | | | | | OOo(Java) | 1.73 |
| 24 | | | | | Netbeans | 2.06 |
| *Average* | | *1.42* | | *1.39* | | *2.02* |

The data in Table 5 show that the 10 projects in *C* language, the 14 projects in *C++*, and the 24 projects in *Java* all have different *CV*s. Therefore, we know that *open source projects have different naming consistency*.

Moreover, after examining the *CV* value in Table 5, it is obvious that the values of projects in *Java* are, on the whole, higher than those of projects in *C* and *C++*. The average *CV* of *C*, *C++* and *Java* projects are 1.42, 1.39, and 2.02 respectively. Also, three curves in Fig. 3 show that fluctuating margin of *Java* projects is minimum. Its *CV* values hold on the 2.00 level from beginning to the end. In contrast, the curve of *C* and *C++* projects have greater fluctuations, especially of *C++*, whose biggest *CV* is close to 2.00 while the smallest one is lower than 1.00. Therefore, *Java* projects have the highest consistency in naming conventions, *C* projects follow and *C++* projects stand last.

## V. REMARKS AND SUGGESTIONS

Identifiers are clearly important to comprehending the concepts in a program [14] so that the research on the adoption of identifier naming is of great importance. While lots of proprietary software projects are running on computers globally, open source projects are showing great vitality in computer world. Therefore, to uncover the current adoption status of identifier naming, 48 open source projects are involved in this study.

This investigation mainly focuses two issues: *popularity of identifier naming*, and *consistency of identifier naming*. As for the former, *Camel* and *Underline* conventions have highest popularity in *C* projects, *Camel* and *Pascal* are the highest popular in *Java* projects, and *Camel* and *Underline* are top two in *C++* projects. With regard to the latter, *Java* projects stand at the highest position, *C* projects follow it and *C++* projects are left behind. Coefficient of variation presents distribution degree of naming conventions in every project in detail. By and large, correlation coefficient of *Java* projects is the greatest and its curve is the most stable. *C++* projects have a larger fluctuation.

In this study, we do not mean to rank the three programming languages *C*, *C++* and *Java*. Meanwhile, we do not judge which is the best identifier naming convention either. Every programming language is a miracle of human intelligence in the fields of computer science and mathematics. Also, every identifier naming convention has an interesting history. As we know, *C* programming language was invented the earliest, *C++* followed, and now *Java* becomes one of the most popular languages. Generally, today's *C++* programmers are accustomed to identifier naming habits they learned from *C* programs while they are building up some new naming convention habits from *Java* language. That may be why naming conventions in *C++* projects disperse and naming conventions in *Java* projects have better performance. From the performance of *C*, *C++* and *Java*, it is possible to envisage that, as to identifier naming conventions, mainstream programming languages are interacting and learning from each other. Identifier naming emerges a convergence trend.

Based on the above research, our recommendations to the managers of software organizations, whether you are developing open source projects or proprietary ones, include: (1) require your software programmers to adopt the most popular identifier naming conventions such as *Camel*, *Pascal* and *Capital* so that it will greatly facilitate the cooperation with other software organizations or within your own organization; (2) monitor the consistency of identifier naming by your programmers when you are developing specific software project because consistent identifier naming may decrease maintenance cost and reduce the risk from the mobility of talents.

Similarly, to the educators of educational institutes, the recommendations embrace: (1) develop your students' habit of utilizing the most popular identifier naming conventions mentioned above and inform them that it will be helpful for them to get a competitive position in software industry in the future; (2) train your students to enhance their cognition of consistent identifier naming and be well prepared to develop or manage large-scale software projects.

**Appendix A** Language Popularity Ranking by TIOBE Software in April 2013 (Top 5)

| *Position Apr 2013* | *Position Apr 2012* | *Delta in Position* | *Programming Language* | *Ratings Apr 2013* | *Delta Apr 2012* | *Status* |
|---|---|---|---|---|---|---|
| 1 | 1 | = | C | 17.862% | +0.31% | A |
| 2 | 2 | = | Java | 17.681% | +0.65% | A |
| 3 | 3 | = | C++ | 9.714% | +0.82% | A |
| 4 | 4 | = | Objective-C | 9.598% | +1.36% | A |
| 5 | 5 | = | C# | 6.150% | -1.20% | A |

Note: status "A" means this language is considered to be mainstream language

**Appendix B** Prefixes in Hungarian naming convention

| *Prefix* | *Type* | *Prefix* | *Type* | *Prefix* | *Type* |
|---|---|---|---|---|---|
| a | Array | cy | Short Int | m_ | Member |
| b | Boolean | dw | Double Word | n | Short Int |
| by | Byte | fn | function | np | Near Pointer |
| c | Char | h | Handle | p | Pointer |
| cb | Char Byte | i | integer | s | String |
| cr | Color Ref | l | Long Int | sz | String with zero end |
| cx | Short Int | lp | Long Pointer | w | Word |

**Appendix C** Considered files types of three programming languages

| *Language* | *File types* |
|---|---|
| C | *.c, *.h |
| C++ | *.c, *.h, *.cpp, *.hpp |
| Java | *.java |

**Appendix D** Structure and example rows in database table *Result*

| *Project* | *Version* | *Language* | *Pascal* | … | *Total No. of ID* | *Total LOC* | *Total No. of Files* |
|---|---|---|---|---|---|---|---|
| checkstyle | 5.1 | Java | 987 | … | 9328 | 51265 | 767 |
| codeblocks | 8.02 | C++ | 8661 | … | 32738 | 277623 | 1356 |
| MingGW | 5.1.6 | C | 849 | … | 73179 | 1069848 | 5068 |
| ... | ... | ... | ... | ... | ... | ... | ... |

**Appendix E** Naming convention usage in 10 projects written in **C** language (sorted by LOC)

| Project name | Version | Pascal | Camel | Hungarian | Underline | Capital | TotalID | TotalLOC |
|---|---|---|---|---|---|---|---|---|
| haproxy | 1.4.6 | 0 | 137 | 0 | 87 | 0 | 276 | 3,037 |
| cherokee | 1.0.1 | 20 | 2,186 | 36 | 1,346 | 36 | 3,926 | 59,009 |
| zabbix | 1.8.2 | 12 | 3,581 | 46 | 2,013 | 1,221 | 7,485 | 73,863 |
| nginx | 0.8.35 | 0 | 3,743 | 0 | 2,045 | 15 | 6,349 | 78,256 |
| httpd | 2.2.15 | 86 | 10,861 | 303 | 6,839 | 260 | 20,187 | 220,287 |
| net-snmp | 5.5 | 169 | 8,568 | 121 | 4,623 | 455 | 15,194 | 240,010 |
| subversion | 1.6.11 | 83 | 9,353 | 42 | 7,762 | 115 | 23,801 | 551,513 |
| MingGW | 5.1.6 | 849 | 36,587 | 180 | 23,372 | 1,780 | 73,179 | 1,069,848 |
| gcc | 4.4.2 | 18,872 | 83,598 | 190 | 33,309 | 7,800 | 180,413 | 2,171,562 |
| linux | 2.6.34 | 9,619 | 376,518 | 4,154 | 421,534 | 34,979 | 931,130 | 7,949,745 |

**Appendix F** Naming convention usage in 14 projects written in **C++** language (sorted by LOC)

| Project name | Version | Pascal | Camel | Hungarian | Underline | Capital | TotalID | TotalLOC |
|---|---|---|---|---|---|---|---|---|
| UnitTest++ | 1.4 | 165 | 178 | 0 | 11 | 1 | 387 | 3,844 |
| sockets | 2.3.9.2 | 540 | 647 | 19 | 441 | 117 | 2,078 | 19,761 |
| GigaBASE | 3.7.7 | 381 | 3,813 | 156 | 648 | 156 | 5,766 | 62,522 |
| ekiga | 3.2.6 | 326 | 2,050 | 17 | 1,393 | 201 | 4,667 | 71,567 |
| LiteSQL | 0.3.8 | 346 | 3,238 | 923 | 489 | 121 | 6,196 | 80,059 |
| PorkerTH | 0.7.1 | 1,203 | 6,432 | 19 | 1,008 | 107 | 10,251 | 111,899 |
| Nestopia | 1.40 | 5,491 | 4,836 | 160 | 85 | 2,120 | 13,511 | 115,879 |
| codeblocks | 8.02 | 8,661 | 14,740 | 891 | 2,074 | 1,262 | 32,738 | 277,623 |
| MySQL | 5.1.45 | 3,772 | 25,795 | 344 | 18,261 | 3,886 | 58,306 | 783,228 |
| CLucene | 0.9.21 | 3,729 | 33,845 | 223 | 11,427 | 1,804 | 92,271 | 823,721 |
| qt | 4.6.2 | 19,300 | 170,731 | 4,051 | 16,285 | 6,568 | 257,611 | 2,374,217 |
| firefox | 3.6 | 7,364 | 139,569 | 4,960 | 17,371 | 12,331 | 205,286 | 2,633,979 |
| CodeLite | 2.5.3 | 68,097 | 192,343 | 21,615 | 38,509 | 10,088 | 396,898 | 3,939,861 |
| Shareaza | 2.5.3.0 | 102,014 | 162,386 | 264,957 | 29,223 | 21,920 | 676,267 | 6,383,063 |

**Appendix G** Naming convention usage in 24 projects written in **Java** language (sorted by LOC)

| Project name | Version | Pascal | Camel | Hungarian | Underline | Capital | TotalID | TotalLOC |
|---|---|---|---|---|---|---|---|---|
| tiles | 2.1.4 | 320 | 2,466 | 1 | 0 | 98 | 2,986 | 17,557 |
| ftpserver | 1.0.4 | 241 | 2,819 | 8 | 0 | 402 | 3,553 | 19,639 |
| mina | 1.1.7 | 363 | 3,239 | 8 | 7 | 201 | 3,984 | 21,820 |
| velocity | 1.6.3 | 403 | 3,751 | 5 | 32 | 399 | 5,162 | 36,860 |
| FreeMind | 0.9.0 RC7 | 461 | 6,574 | 56 | 12 | 287 | 7,943 | 39,179 |
| freemarker | 2.3.16 | 477 | 4,172 | 22 | 40 | 470 | 5,830 | 43,048 |
| checkstyle | 5.1 | 987 | 7,171 | 22 | 22 | 469 | 9,328 | 51,265 |
| DependencyFinder | 1.2.1B3 | 651 | 7,935 | 4 | 60 | 445 | 10,312 | 58,991 |
| pmd | 4.2.5 | 805 | 6,573 | 22 | 107 | 925 | 9,901 | 60,617 |
| struts2 | 2.1.6 | 1,151 | 11,967 | 11 | 11 | 479 | 14,078 | 62,992 |
| dwr | 3.0.0.116 | 833 | 8,859 | 64 | 17 | 672 | 10,730 | 77,188 |
| hibernate | 3.0 RC1 | 1,089 | 14,823 | 5 | 3 | 476 | 17,423 | 94,930 |
| findbugs | 1.3.9 | 1,237 | 18,857 | 60 | 42 | 1,146 | 22,607 | 110,473 |
| ant | 1.8.1 | 1,367 | 22,050 | 57 | 35 | 1,949 | 27,545 | 126,218 |
| JFreeChart | 1.0.13 | 967 | 17,224 | 36 | 0 | 750 | 22,116 | 127,816 |
| jruby | 1.4.0 | 1,287 | 14,413 | 75 | 166 | 1,118 | 18,510 | 149,580 |
| tomcat | 6.0.26 | 1,470 | 24,985 | 124 | 64 | 1,815 | 30,195 | 168,819 |

| JasperReports | 3.7.3 | 746 | 23,437 | 33 | 0 | 2,225 | 28,663 | 190,634 |
| axis2 | 1.4.1 | 2,368 | 43,046 | 56 | 450 | 3,173 | 54,221 | 288,287 |
| spring-framework | 3.0.0 | 5,614 | 47,255 | 16 | 11 | 1,832 | 56,235 | 318,107 |
| openjpa | 2.0.0 | 3,953 | 52,616 | 194 | 55 | 2,046 | 67,351 | 353,442 |
| JEdit | 4.3.2 | 4,012 | 43,186 | 872 | 1,182 | 3,604 | 61,392 | 368,669 |
| OpenOffice(Java) | 3.2.0 | 6,073 | 53,800 | 2,383 | 702 | 3,920 | 77,575 | 408,254 |
| Netbeans | 6.8 | 41,962 | 584,997 | 2,217 | 2,162 | 43,410 | 726,525 | 3,898,230 |